# Dimensional crossover of a boson gas in multilayers

P. Salas[1,2], F.J. Sevilla[2], M. Fortes[2], M. de Llano[3], A. Camacho[4], and M.A. Solís[2]

[1]Posgrado en Ciencias e Ingeniería de Materiales, UNAM,
Apdo. Postal 70-360, 04510 México, D.F., MEXICO
[2]Instituto de Física, UNAM, Apdo. Postal 20-364, 01000 México, D.F., MEXICO
[3]Instituto de Investigaciones en Materiales, UNAM,
Apdo. Postal 70-360, 04510 México, D.F., MEXICO
[4]Facultad de Ingeniería, UNAM, 04510 México, D.F., MEXICO

We obtain the thermodynamic properties for a non-interacting Bose gas constrained on multilayers modeled by a periodic Kronig-Penney delta potential in one direction and allowed to be free in the other two directions. We report Bose-Einstein condensation (BEC) critical temperatures, chemical potential, internal energy, specific heat, and entropy for different values of a dimensionless impenetrability $P \geqslant 0$ between layers. The BEC critical temperature $T_c$ coincides with the ideal gas BEC critical temperature $T_0$ when $P = 0$ and rapidly goes to zero as $P$ increases to infinity for any finite interlayer separation. The specific heat $C_V$ vs $T$ for finite $P$ and plane separation $a$ exhibits one minimum and one or two maxima in addition to the BEC, for temperatures larger than $T_c$ which highlights the effects due to particle confinement. Then we discuss a distinctive dimensional crossover of the system through the specific heat behavior driven by the magnitude of $P$. For $T < T_c$ the crossover is revealed by the change in the slope of $\log C_V(T)$ and when $T > T_c$, it is evidenced by a broad minimum in $C_V(T)$.

## I. INTRODUCTION

Non-relativistic composite bosons in layered structures are found as molecular electron pairs in cuprate superconductors [1], alkaline atoms in optical lattices [2], excitons in multilayered semiconductors [3], exciton polaritons in microcavities [4] or simply as atoms of helium four adsorbed on graphite [5] or any another substrate [6, 7]. These systems with planar symmetry have been described in [8, 9] and several periodic potentials have been used such as the sinusoidal [10] and the biparabolic [11] with good results only in the low particle energy limit, or in the tight-binding approximation [12]. However, to analyze structural effects like particle trapping between the planes or quasi-2D behavior when plane separation is of the order of half the thermal wavelength, it is necessary to consider a much wider temperature region in which a large number of energy bands needs to be included in the calculation of the system thermodynamic properties.

Recent experiments with ultra-cold atoms particularly in optical lattices, provide a test not only to probe new physics at very low temperatures but also for interacting many-body models of condensed matter (the so called Quantum Simulators [13]). Although this is the current trend in the field, experimental and theoretical studies on non-interacting many-body systems [10, 14] are still being considered due to the possibility of tuning off the strength of interactions between particles in experimental situations.

In Ref. [15] it was shown that a layered structure with plane separation $a$ is revealed in the specific heat behavior as a function of temperature. For closely separated planes ($a \lesssim \lambda_0 \equiv h/\sqrt{2\pi m k_B T_0}$), a minimum at $T \equiv T_{min}$ whose corresponding thermal wavelength scales with $a$ as $\lambda \simeq 2a$ is found in the $C_V$ vs $T$ curves, where $T_0$ is the ideal gas BEC critical temperature. For $a \gtrsim \lambda_0$ the minimum disappears as a well-defined bump develops while the sharpness of the transition smooths out. This bump occurs at $T_{max} > T_c$ and in the limit of $a \to \infty$ both, $T_c$ and $T_{max}$ tend to $T_0$ and the bump evolves into the well known BEC peak.

In this paper we show that there is a marked dimensional crossover from 3D to quasi-2D as $P$ [16] is increased over a range of thermal energies around $\pi \hbar^2 / 2ma^2$. Here we elucidate some aspects of this feature. Even when interactions between bosons are neglected, it is known that the inhomogeneity induced by a trapping potential such as a harmonic trap [17] deforms the free particle density-of-states (DOS) thus giving rise to a finite, non-vanishing, critical temperature in a 2D system. In a weakly-interacting Bose gas the trapping potential suppresses long-range thermal fluctuations that would otherwise destroy the condensate [18].

The paper unfolds as follows. In §II we describe the model consisting of a boson gas initially in an infinitely large box where we introduce layers of null width and finite impenetrability separated periodically by a distance $a$. The layers are introduced via the Kronig-Penney (KP) potential [19] only in the $z$-direction and in the limit where the KP square potential barriers become repulsive delta potentials. Using the energy dispersion relations in this model, we derive the number equation and the equation for the grand potential to calculate the critical temperature and the condensate fraction. From the grand potential we also obtain the internal energy, specific heat, chemical potential and entropy, which are then compared with the case where there are no layers, i.e., with the properties of the infinite ideal boson gas. In §III the density of states is obtained and a model of it is proposed to exhibit the 3D to quasi-2D dimensional crossover. In §IV results are discussed and conclusions given.

## II. THERMODYNAMIC PROPERTIES

We consider a system of $N$ non-interacting bosons within layers of separation $a$ which we model as a periodic array of delta potentials of strength $\Lambda$ in the $z$-direction, $\sum_{n=-\infty}^{\infty} \Lambda \delta(z - na)$. They are free in the other two directions (see Fig. 1). This $z$-directional model is a version of the Kronig-Penney (KP) potential known as the "Dirac comb" potential . The Schrödinger equation for any boson of mass $m$ is separable in $x$, $y$ and $z$-directions so that the single-particle energy as a function of the momentum $\mathbf{k} = (k_x, k_y, k_z)$ is $\varepsilon_{\mathbf{k}} = \varepsilon_{k_x} + \varepsilon_{k_y} + \varepsilon_{k_z}$, where

$$\varepsilon_{k_x, k_y} = \frac{\hbar^2 k_{x,y}^2}{2m} \qquad (1)$$

with $k_{x,y} = 2\pi n_{x,y}/L$ and $n_{x,y} = 0, \pm 1, \pm 2, ...$, i.e., in the $x$ and $y$-directions particles are free and periodic boundary conditions in a box of size $L$ are imposed. The energies $\varepsilon_{k_z}$ in the $z$-direction are obtainable from

$$(P/\alpha a)\sin(\alpha a) + \cos(\alpha a) = \cos(k_z a) \qquad (2)$$

with $\alpha^2 \equiv 2m\varepsilon_{k_z}/\hbar^2$ and we have defined the dimensionless constant $P \equiv m\Lambda a/\hbar^2$ as the ratio of the competing energy scales, $\Lambda/a$ and $\hbar^2/ma^2$. Here $P$ measures the layer impenetrability whereby $P \to 0$ signifies the trivial free-particle dispersion energy in the $z$-direction $\varepsilon_{k_z} \to \hbar^2 k_z^2 / 2m$. By contrast, $P \to \infty$ implies $\sin(\alpha a) \to 0$ which corresponds to confining bosons inside a semi-infinite slab of width $a$ and lateral infinite extent as has been extensively discussed in the literature (see [21] and references therein). For small particle energy, i.e. $\varepsilon_{k_z} \ll \hbar^2/2ma^2$, (2) becomes

$$\varepsilon_{k_z} \simeq \varepsilon_0 + \frac{\hbar^2}{Ma^2}(1 - \cos k_z a) \qquad (3)$$

where $\varepsilon_0$ and $M$ are the $P$-dependent ground-state energy and the effective boson mass in the $z$ direction, respectively. Here, $\varepsilon_0$ satisfies the equation (2) for $k_z \to 0$, namely $(P/\alpha_0 a)\sin(\alpha_0 a) + \cos(\alpha_0 a) = 1$ where $\varepsilon_0 \equiv \hbar^2 \alpha_0^2/2m$ and $M \equiv [[\sin \alpha_0 a - (P+1)(\cos \alpha_0 a)/\alpha_0 a]/\alpha_0 a]$. Expression (3) has been used [8, 9] to describe the dispersion relation in periodic arrays such as the $CuO_2$ planes in cuprate superconductors. However, these studies were limited to very low energies since only the first energy band was considered.

Calculation of thermodynamic properties requires integration, in the thermodynamic limit, over all allowed wavevectors $\mathbf{k}$. Integrals over $k_x$ and $k_y$ are straightforward but for the integration over $k_z$, the values of $\varepsilon_{k_z}$ are required from the solution of (2). This is discussed extensively, e.g., in Refs. [22, 23], where $(\alpha a)^2$ as a function of $k_z a$ exhibits allowed and forbidden energy bands. The integrals over $k_z$ can be evaluated as the sum over allowed bands of integrals over the first half-Brillouin zone ($0 < k_z a < \pi$). In other words, for any function $f(\varepsilon_z)$,

$$\int_{-\infty}^{\infty} dk_z f(\varepsilon_{k_z}) = 2 \sum_{j=1}^{\infty} \int_0^{\pi/a} dk_z f(\varepsilon_{k_z j})$$

where $\varepsilon_{k_z j}$ denotes the energy in the $j$-th band.

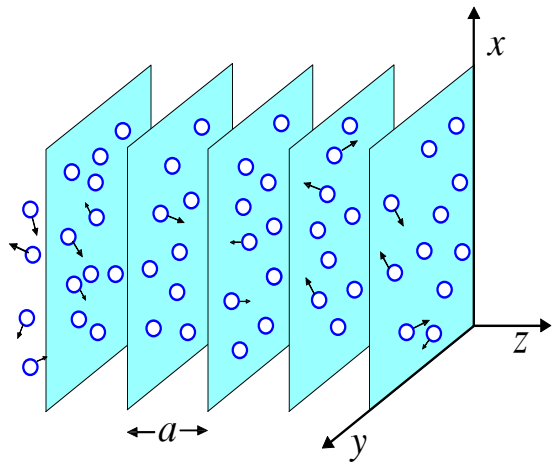

FIG. 1: (Color online) Bosons among very thin planes of variable penetrability and stacked in the $z$-direction.

Thermodynamic properties are then obtained from the grand potential $\Omega(T, V, \mu)$ for a boson gas [24] in a volume $V \equiv L^3$, namely

$$
\begin{aligned}
\Omega(T, V, \mu) &= U - TS - \mu N \qquad (4) \\
&= \Omega_0 + k_B T \sum_{\mathbf{k} \neq 0} \ln[1 - \exp\{-\beta(\varepsilon_{k_i} - \mu)\}]
\end{aligned}
$$

where $U$ is the internal energy, $S$ the entropy, $\beta \equiv 1/k_B T$ and $\mu$ is the chemical potential. Here we have explicitly separated the term $\Omega_0 \equiv k_B T \ln[1 - \exp\{-\beta(\varepsilon_0 - \mu)\}]$ corresponding to the $\mathbf{k} = 0$ ground-state contribution to the grand potential. After expanding the logarithm $\ln(1 + x)$ into its power series $-\sum_{l=1}^{\infty}(-x)^l/l$ valid for $|x| < 1$, substituting $\varepsilon_{k_i}$ and reordering, (4) becomes

$$
\begin{aligned}
\Omega(T, V, \mu) &= \Omega_0 - k_B T \sum_{l=1}^{\infty} \frac{\exp \beta\mu l}{l} \times \\
&\sum_{\mathbf{k} \neq 0} \exp\{-\beta l\left[(\hbar^2/2m)(k_x^2 + k_y^2) + \varepsilon_{k_z}\right]\}.
\end{aligned}
$$

In the continuous limit where $\hbar^2/mL^2 \ll k_B T$, i.e., when level spacing is negligible compared to thermal energy, the summations over $\mathbf{k}$ can be approximated by integrals, namely $\sum_{\mathbf{k}} \longrightarrow (L/2\pi)^3 \int d^3\mathbf{k}$. Thus

$$\Omega(T,V,\mu) = \Omega_0 - k_B T \left(\frac{L}{2\pi}\right)^3 \sum_{l=1}^{\infty} \frac{\exp \beta l \mu}{l} \times$$

$$\int_{-\infty}^{\infty} dk_x \exp\{-\beta l(\hbar^2/2m)k_x^2\} \times$$

$$\int_{-\infty}^{\infty} dk_y \ \exp\{-\beta l(\hbar^2/2m)k_y^2\} \int_{-\infty}^{\infty} dk_z \exp\{-\beta l \varepsilon_{k_z}\}.$$

The integrals over $k_x$, $k_y$ are elementary so that

$$\Omega(T,V,\mu) =$$
$$\Omega_0 - \frac{1}{\beta^2} \frac{L^3 m}{(2\pi)^2 \hbar^2} \int_{-\infty}^{\infty} dk_z \sum_{l=1}^{\infty} \frac{\exp \beta l(\mu - \varepsilon_{k_z})}{l^2}. \quad (5)$$

The infinite sum is expressible in terms of Bose functions [24] $g_\sigma(t) = \sum_{l=1}^{\infty} t^l/l^\sigma$. Combining $g_\sigma$ and (5) leaves

$$\Omega(T,V,\mu) = \Omega_0 - \frac{1}{\beta^2} \frac{L^3 m}{(2\pi)^2 \hbar^2} \int_{-\infty}^{\infty} dk_z \times \quad (6)$$

$$g_2(\exp\{\beta(\mu - \varepsilon_{k_z})\}). \quad (7)$$

From (6) it is possible to find the thermodynamic properties for a monatomic gas using the well-known relation

$$d\Omega = -SdT - pdV - Nd\mu. \quad (8)$$

In this representation the grand potential $\Omega(T,V,\mu) = -pV$ is the fundamental relation leading to all the thermodynamic properties of the system, via [24]

$$N = -\left(\frac{\partial \Omega}{\partial \mu}\right)_{T,V}, \quad S = -\left(\frac{\partial \Omega}{\partial T}\right)_{V,\mu}$$

$$p = -\left(\frac{\partial \Omega}{\partial V}\right)_{T,\mu} = -\frac{\Omega}{V}. \quad (9)$$

The internal energy and specific heat are

$$U(T,V) = -k_B T^2 \left[\frac{\partial}{\partial T}\left(\frac{\Omega}{k_B T}\right)\right]_{V,z}$$

$$\text{and } C_V = \left[\frac{\partial}{\partial T} U(T,V)\right]_{N,V}, \quad (10)$$

where $z \equiv \exp(\beta\mu)$ is the fugacity.

## A. Critical temperature

The equation for particle number $N$ is obtained from the first equation of (9) and the grand potential (6) to yield

$$N = \frac{1}{\exp\{\beta(\varepsilon_0 - \mu)\} - 1} - \quad (11)$$
$$\frac{L^3 T}{(2\pi)^2 T_0 \gamma a^2} \int_0^{\infty} dk_z \ln\left(1 - \exp\{-\beta(\varepsilon_{k_z} - \mu)\}\right)$$
$$\equiv N_0(T) + N_e(T)$$

where we have introduced the dimensionless parameter $\gamma \equiv \hbar^2/2ma^2 k_B T_0 = (1/4\pi)(\lambda_0/a)^2$, $T_0$ being the critical temperature of an ideal boson gas in an infinite box given by $T_0 = 2\pi\hbar^2 n_B^{2/3}/mk_B\zeta(3/2)^{2/3} \simeq 3.31\hbar^2 n_B^{2/3}/mk_B$ with $n_B \equiv N/L^3$ the boson number density. Evidently, $\gamma$ and $n_B$ are related by

$$\gamma^{3/2} = \zeta(3/2)/(4\pi)^{3/2} a^3 n_B \simeq 0.0586/N_a$$

where $N_a$ is the number of bosons contained inside a volume $a^3$. Clearly, the first term in (11) is the number of particles $N_0(T)$ in the condensed state while the second term is the number of bosons $N_e$ in excited states. For $T > T_c$, $N_0(T)$ is negligible compared with $N$ while for $T < T_c$ it becomes a sizeable fraction of $N$. At precisely $T = T_c$, $\mu = \mu_0$, $N_0(T_c) \simeq 0$ and $N \simeq N_e$, with $\mu_0$ the lowest particle energy. Then, at $T = T_c$ (11) is just

$$N = -\frac{L^3}{(2\pi)^3} \frac{2\pi}{a^3} \frac{T_c}{T_0 \gamma} \times$$
$$\int_0^{\infty} a \, dk_z \ln(1 - \exp\{-\beta_c(\varepsilon_{k_z} - \mu_0)\}). \quad (12)$$

Since $\varepsilon_{k_z}$ has a band structure one can split the integral into a sum of contributions from each of the allowed energy bands. In practice, the infinite sum is truncated after, say, $J$ terms once convergence is achieved. The expression for the critical temperature is then

$$-1 = \frac{AT_c}{T_0} \sum_{j=1}^{J} \int_{(j-1)\pi}^{j\pi} d\xi \ln(1 - \exp\{-\beta_c(\varepsilon_{k_z} - \mu_0)\}) \quad (13)$$

where $A \equiv 2(\gamma/\pi)^{1/2}/\zeta(3/2)$, $\xi \equiv ak_z$ and we have reexpressed $N$ in terms of the critical temperature $T_0$. Using $\xi \equiv \eta + (j-1)\pi$ then $d\xi = d\eta$, and when $\xi = (j-1)\pi$ then $\eta = 0$, similarly when $\xi = j\pi$, $\eta = \pi$. Using this definition (13) becomes

$$-1 = A\tilde{T}_c \sum_{j=1}^{J} \int_0^{\pi} d\eta \ln(1 - \exp\{-\tilde{\beta}_c \gamma(\bar{\varepsilon}_{\xi j} - \bar{\mu}_0)\}) \quad (14)$$

if one introduces the dimensionless quantities $\tilde{\beta} \equiv \beta k_B T_0$ and $\tilde{T}_c = T_c/T_0$. Here, $\bar{\varepsilon}_{\xi j}$ and $\bar{\mu}_0$ are the energy and chemical potential below $T_c$, respectively, in units of $\hbar^2/2ma^2$ and are the solutions of (2). Since the lhs of (14) is constant this is an implicit equation to determine $T_c$. In Fig. 2 we show the critical temperature as a function of the parameter $P$ for five values of $\gamma$ which were chosen to span a wide symmetric range of $a$ values around $a = \lambda_0$. For reference, the interval over which empirical HTSC critical temperatures are found [25, 26] is shown between dashed horizontal lines. We note that $\tilde{T}_c$ is a monotonically decreasing function of $P$ but not of $\gamma$ or $\lambda_0/a$ as shown by crossing curves [15] for $\gamma = 1, 10$. The existence of a finite *nonzero* critical temperature for any finite value of $P$ is caused by the $\varepsilon^{1/2}$ behavior of the density of states in 3D, valid for energies around the bottom of the first band (see §III). In the limit $P \to \infty$ we recover a collection of independent slabs of volume $L \times L \times a$ with $L \to \infty$ and finite $a$, such that the $z$-direction particle energies are those associated with Dirichlet boundary conditions. The critical temperature for any of these slabs with boson density $n_B$ vanishes as a consequence of an infrared divergence as occurs in the exact 2D system [27].

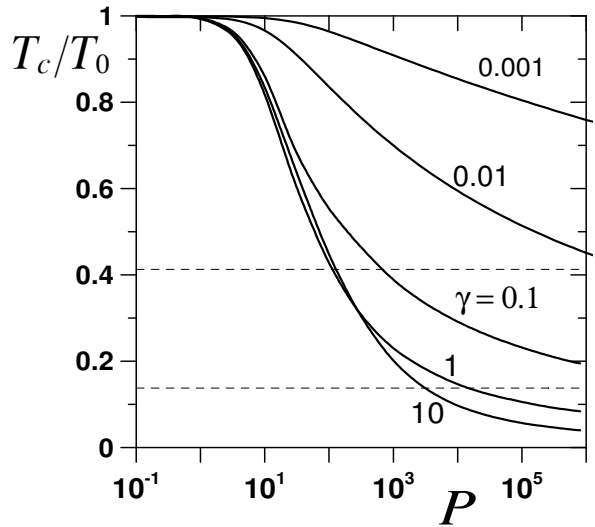

FIG. 2: Critical temperature as a function of the impenetrability parameter $P$ for different values of $\gamma$. Critical temperatures of HTSC lie in the region between the horizontal dashed lines.

### B. Condensate fraction

From (11), the condensate fraction $N_0/N = 1 - N_e/N$ is

$$N_0/N = 1 + \frac{A}{\bar{\beta}} \sum_{j=0}^{J} \int_0^\pi d\xi \ln(1 - \exp\{-\tilde{\beta}\gamma(\bar{\varepsilon}_{\xi j} - \bar{\mu}_0)\}).$$
(15)

In Fig. 3 we compare the condensate fractions of an ideal free Bose gas, i.e. $P = 0$, and of bosons among layers with $P = 10, 10^2, 10^3$ and $10^4$ with $a/\lambda_0 = 0.892$ (upper panel) and $a/\lambda_0 = 0.089$ (lower panel). The different behavior of $N_0/N$ in each case may be explained as follows: the smaller the ratio $a/\lambda_0$, the larger the temperature $T_{min}$ where confinement effects are markedly observed. This is the case for $a/\lambda_0 = 0.089$ ($\gamma = 10$) where $T_{min} \simeq 31.42\,T_0 > T_c$ (see Table I for the respective values of $T_c/T_0$), and therefore we do not expect a large departure from the $P = 0$ case. On the other hand, trapping effects are revealed at low temperatures when $a/\lambda_0 \gtrsim 0.892$ and are more prominent as $P$ increases.

In particular, for $a/\lambda_0 = 0.892$ ($T_{min} \simeq 0.314\,T_0$), the condensate fraction changes from its familiar temperature dependence $1 - (T/T_c)^{3/2}$ for $P = 0$, to $1 - (T/T_c)^{d/2}$ while the effective dimension $2 < d < 3$ is a function of $P$ and $a/\lambda_0$. A fit to this expression leads to $d = 2.842, 2.558, 2.376$ and $2.275$ for $P = 10, 10^2, 10^3$ and $10^4$ respectively as shown in the upper panel of Fig. 3.

TABLE I: Critical temperatures $T_c/T_0$ computed with the values of $P$ and $T_{min}/T_0$, for $\gamma$ used in Figs. 3-8 ($T_c/T_0 = 1$ for $P = 0$).

| $T_{min}/T_0$ | $\gamma$ \ $P$ | 10 | $10^2$ | $10^3$ | $10^4$ |
|---|---|---|---|---|---|
| 0.31416 | 0.1 | 0.859 | 0.566 | 0.39 | 0.292 |
| 31.416 | 10 | 0.834 | 0.449 | 0.203 | 0.098 |

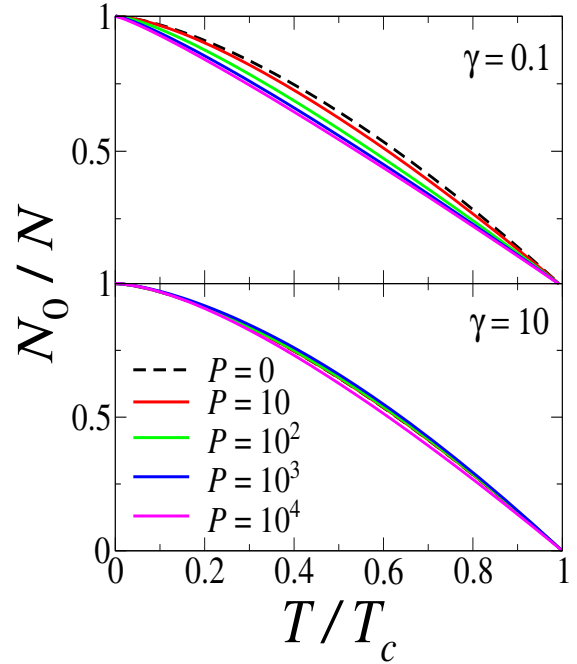

FIG. 3: (Color online) Condensate fraction as a function of $T/T_c$ for different values of the impenetrability parameter $P$. Upper panel is for $a/\lambda_0 = 0.892$; lower panel is for $a/\lambda_0 = 0.089$.

### C. Internal energy

The internal energy $U$ is obtained from (6) and (10) as

$$U(T,V) = \frac{\varepsilon_0}{\exp\beta(\varepsilon_0 - \mu) - 1}$$
$$- \frac{L^3 m}{(2\pi)^2 \hbar^2} \frac{2}{\beta} \int_0^\infty dk_z \varepsilon_{k_z} \ln[1 - \exp\{-\beta(\varepsilon_{k_z} - \mu)\}]$$
$$+ \frac{L^3 m}{(2\pi)^2 \hbar^2} \frac{2}{\beta^2} \int_0^\infty dk_z g_2(\exp\{-\beta(\varepsilon_{k_z} - \mu)\})$$
(16)

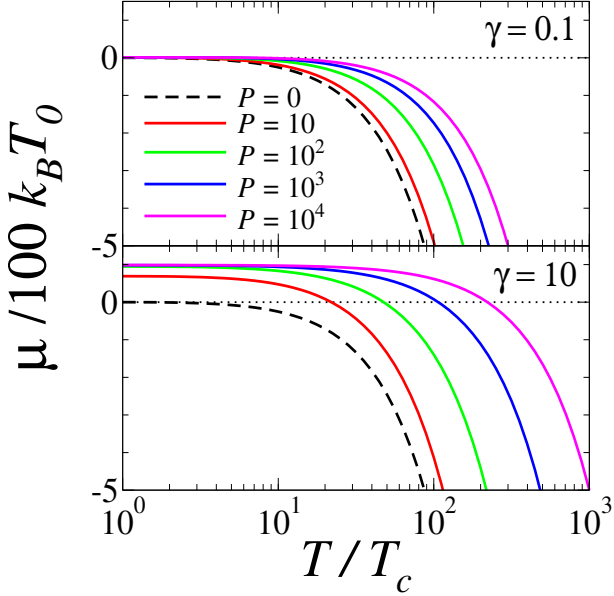

FIG. 4: (Color online) Chemical potential in $k_B T_0$ units as a function of $T/T_c$ for different values of the impenetrability parameter $P$. Upper panel is for $a/\lambda_0 = 0.892$; lower panel is for $a/\lambda_0 = 0.089$.

which can be rewritten as

$$(U - N\varepsilon_0)/Nk_B T = \frac{A}{\tilde{\beta}} \sum_{j=1}^{J} \int_0^\pi d\xi \, g_2(\exp\{-\tilde{\beta}(\tilde{\varepsilon}_{k_{zj}} - \tilde{\mu})\})$$

$$- A \sum_{j=1}^{J} \int_0^\pi d\xi \, (\tilde{\varepsilon}_{k_{zj}} - \tilde{\varepsilon}_0) \ln(1 - \exp\{-\tilde{\beta}(\tilde{\varepsilon}_{k_z} - \tilde{\mu})\})$$

(17)

To find $U$, we rewrite (11) to obtain $\mu(T)$ as

$$1 = N_0/N - \frac{A}{\tilde{\beta}} \int_0^\infty a dk_z \ln(1 - \exp\{-\tilde{\beta}(\tilde{\varepsilon}_{k_z} - \tilde{\mu})\}) \quad (18)$$

from which one can numerically extract $\mu(T)$. In Fig. 4 $\mu(T)$ is shown for different values of $P$ and for two values of $\gamma = 0.1$ and 10 in which the $C_V$ has a markedly different behavior [15]. In the limit $P \to \infty$ and $a/\lambda_0 \ll 1$, when the system is approximately two-dimensional, we expect $\mu(T)$ to vary according to $\varepsilon_0 + k_B T \ln\left[1 - \exp(-\hbar^2 n_{2D}/2\pi m k_B T)\right]$, where $n_{2D}$ is the 2D bosonic number density.

Note that the relation $U = \frac{3}{2} pV$ satisfied by ideal quantum gases at all temperatures is no longer valid in the temperature region where confinement effects appear. In this region we have $U = \chi(T) pV$, where the factor $\chi(T)$ changes smoothly from 3/2 to 1 and goes back again to 3/2. This effect is conspicuously evidenced for $a/\lambda_0 < 1$ as is observed by comparing the bottom panels of Figs. 5 and 6.

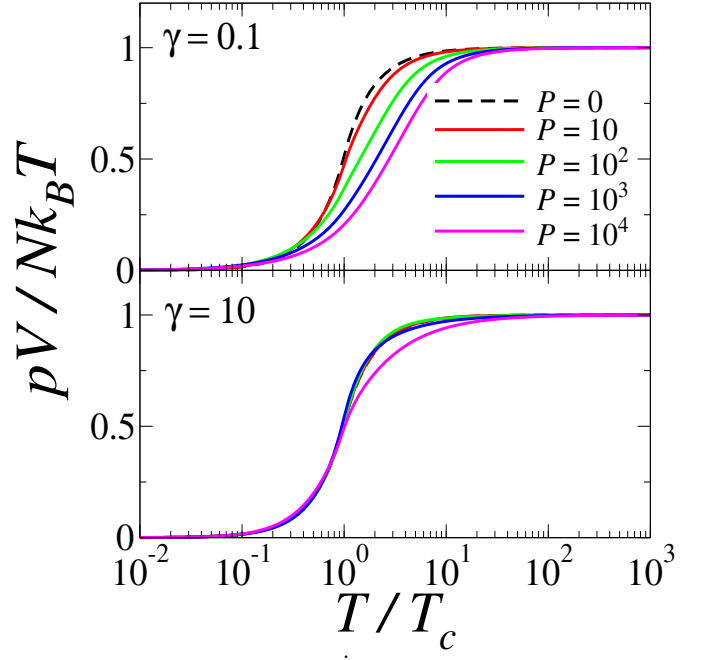

FIG. 5: (Color online) $pV$ in $Nk_B T$ units as a function of $T/T_c$ for different values of the impenetrability parameter $P$. Upper panel is for $a/\lambda_0 = 0.892$; lower panel is for $a/\lambda_0 = 0.089$.

### D. Specific heat

To determine $C_V(T)$ one also needs $\partial \mu(T)/\partial T$. For $T < T_c$, $\mu = \mu_0$ a constant so that $\partial \mu_0/\partial T = 0$. For $T > T_c$, $N_0/N \simeq 0$ and taking the derivative of the last equation with respect to $T$ gives

$$-\frac{d\tilde{\beta}}{dT} = A \int_0^\infty a dk_z \frac{\exp\{-\tilde{\beta}(\tilde{\varepsilon}_{k_z} - \tilde{\mu})\}}{1 - \exp\{-\tilde{\beta}(\tilde{\varepsilon}_{k_z} - \tilde{\mu})\}} \times$$

$$[(\tilde{\varepsilon}_{k_z} - \tilde{\mu})(\partial \tilde{\beta}/\partial T) - \tilde{\beta}(\partial \tilde{\mu}/\partial T)]$$

which after some algebra leads to

$$\frac{1 + A \int_0^\infty a dk_z (\tilde{\varepsilon}_{k_z} - \tilde{\mu})/(\exp\{\tilde{\beta}(\tilde{\varepsilon}_{k_z} - \tilde{\mu})\} - 1)}{A \int_0^\infty a dk_z/(\exp\{\tilde{\beta}(\tilde{\varepsilon}_{k_z} - \tilde{\mu})\} - 1)}$$

$$= -T \frac{\partial \tilde{\mu}}{\partial T}. \quad (19)$$

Taking the derivative of (17), and using (18) and (19) for the chemical potential $\mu$ and its derivative, the specific heat is as follows

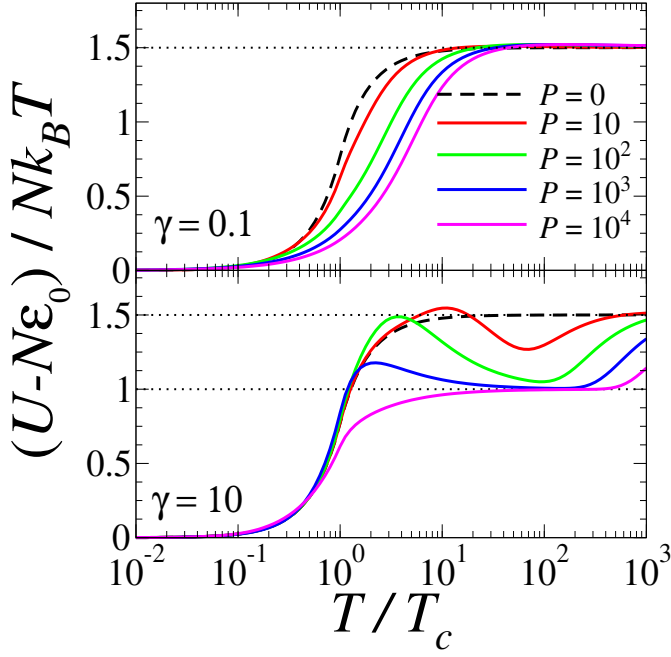

FIG. 6: (Color online) Internal energy in $Nk_BT$ units as a function of $T/T_c$ for different values of the impenetrability parameter $P$. Upper panel is for $a/\lambda_0 = 0.892$; lower panel is for $a/\lambda_0 = 0.089$.

$$\frac{C_V}{Nk_B} = \frac{U - N\varepsilon_0}{Nk_BT} - A\tilde{\mu}_0\tilde{\beta}\int_0^\infty a\,dk_z\frac{(\tilde{\varepsilon}_{k_z} - \tilde{\mu}) + T(\partial\tilde{\mu}/\partial T)}{\exp\{\tilde{\beta}(\tilde{\varepsilon}_{k_z} - \tilde{\mu})\} - 1} + \frac{A}{\tilde{\beta}}\int_0^\infty a\,dk_z\,g_2(\exp\{-\tilde{\beta}(\tilde{\varepsilon}_{k_z} - \tilde{\mu})\})$$
$$- A\int_0^\infty a\,dk_z\,\ln(1 - \exp\{-\tilde{\beta}(\tilde{\varepsilon}_{k_z} - \tilde{\mu})\})\left[(\tilde{\varepsilon}_{k_z} - \tilde{\mu}) + T(\partial\tilde{\mu}/\partial T)\right] + A\tilde{\beta}\int_0^\infty a\,dk_z\,\tilde{\varepsilon}_{k_z}\frac{[(\tilde{\varepsilon}_{k_z} - \tilde{\mu}) + T(\partial\tilde{\mu}/\partial T)]}{\exp\{\tilde{\beta}(\tilde{\varepsilon}_{k_z} - \tilde{\mu})\} - 1} \quad (20)$$

where we used $z[dg_2(z)/dz] = g_1(z) = -\ln(1-z)$. Transforming integrals into sums over allowed bands and after

some algebra gives

$$\frac{C_V}{Nk_B} = \frac{U - N\varepsilon_0}{Nk_BT} + \frac{A}{\tilde{\beta}}\sum_{j=1}^J\int_0^\pi d\xi\,g_2(\exp\{-\tilde{\beta}(\tilde{\varepsilon}_{k_z} - \tilde{\mu})\}) - A\sum_{j=1}^J\int_0^\pi d\xi\,\ln(1 - \exp\{-\tilde{\beta}(\tilde{\varepsilon}_{k_z} - \tilde{\mu})\})\left[(\tilde{\varepsilon}_{k_z} - \tilde{\mu}) + \right.$$
$$T(\partial\tilde{\mu}/\partial T)] + A\tilde{\beta}\sum_{j=1}^J\int_0^\pi d\xi\,(\tilde{\varepsilon}_{k_z} - \tilde{\mu}_0)\frac{[(\tilde{\varepsilon}_{k_z} - \tilde{\mu}) + T(\partial\tilde{\mu}/\partial T)]}{\exp\{\tilde{\beta}(\tilde{\varepsilon}_{k_z} - \tilde{\mu})\} - 1}. \quad (21)$$

As before, for $T < T_c$ the chemical potential $\mu = \mu_0$ is constant and $\partial\tilde{\mu}/\partial T = 0$ so that the specific heat be-

comes

$$\frac{C_V}{Nk_B} = \frac{U - N\varepsilon_0}{Nk_BT} + \frac{A}{\tilde{\beta}}\sum_{j=1}^J\int_0^\pi d\xi\,g_2(\exp\{-\tilde{\beta}(\tilde{\varepsilon}_{k_z} - \tilde{\mu}_0)\})$$
$$- A\sum_{j=1}^J\int_0^\pi d\xi\,\ln(1 - \exp\{-\tilde{\beta}(\tilde{\varepsilon}_{k_z} - \tilde{\mu})\})(\tilde{\varepsilon}_{k_z} - \tilde{\mu}_0)]$$
$$+ A\tilde{\beta}\sum_{j=1}^J\int_0^\pi d\xi\,\frac{(\tilde{\varepsilon}_{k_z} - \tilde{\mu}_0)^2}{\exp\{\tilde{\beta}(\tilde{\varepsilon}_{k_z} - \tilde{\mu}_0)\} - 1}, \quad T < T_c. \quad (22)$$

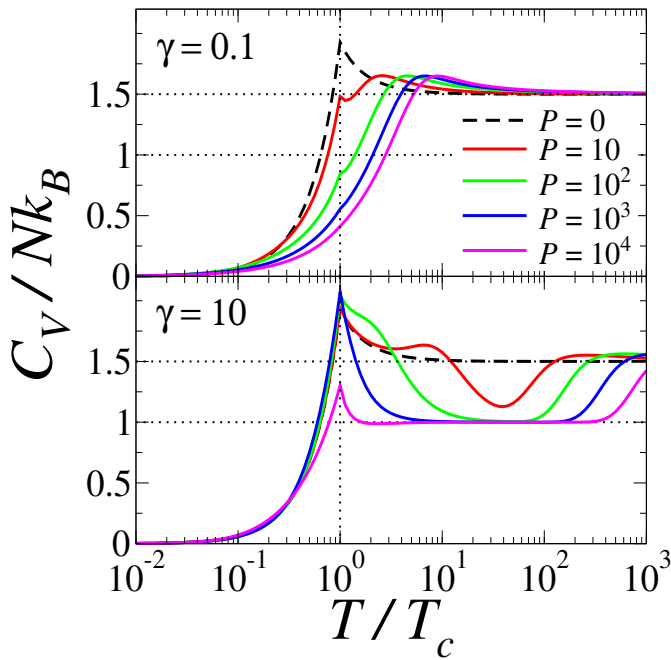

FIG. 7: (Color online) Specific heat in $Nk_B$ units as a function of $T/T_c$, for different values of the impenetrability parameter $P$. Upper panel is for $a/\lambda_0 = 0.892$; lower panel is for $a/\lambda_0 = 0.089$. Horizontal dotted lines denote the classical values for two- and three-dimensional IBG systems.

The effects on the temperature dependence of specific heat when $P$ is increased depends on whether $a \gtrsim \lambda_0$ (top panel in Fig. 7) or $a \lesssim \lambda_0$ (bottom panel in Fig. 7) just as happens for the condensate fraction in §II B. Firstly, plane separations are revealed in the specific heat as a minimum in the $C_V$ vs $T$ curves when $a \lesssim \lambda_0$ (or as a crossover from $T^{3/2}$ to $T$ in the low-temperature dependence of $C_V$ when $a \gtrsim \lambda_0$). Secondly, such characteristics mark an effective 2D behavior of the system on a range of temperatures that becomes wider as $P$ is increased (one can note such behavior in the bottom panel of Fig. 7 where the characteristic $C_V/Nk_B \simeq 1$ in 2D is revealed on a wider range of temperatures as $P$ is increased).

For the $a \gtrsim \lambda_0$ case, two distinctive features in the specific heat can be observed. The first one is characterized by the appearance of a maximum at $T_{max}$ immediately after the BEC $T_c$. Since any property related to the condensate should manifest itself at temperatures lower than $T_c$, this maximum must be unrelated to BEC as is suggested by the fact that it persists even for $P = \infty$ when no BEC exists at all. This conclusion contrasts to the one given in [28–30] where the system studied corresponds to ours in the $P \to \infty$ limit and the maximum in $C_v$ has been used as a criteria that marks the onset of BEC in finite systems.

The second feature contrasts with the 3D case in that the discontinuity of $\partial C_V/\partial T$ at $T_c$, (being positive at $T_c^-$ and negative at $T_c^+$), is reduced as $P$ is increased. Indeed, for large enough values of $P$ the sign of $\partial C_V/\partial T|_{T_c^+}$ changes from negative to positive, thus smoothing the "cusp" of $C_V$ at $T_c$. This same behavior is qualitatively observed when the dimensionality of the IBG is continuously reduced from 3D to 2D [31]. Indeed, a clear signature of 2D behavior is found in the linear temperature dependence of $C_V$ as $T \to T_c$ for $P = 10^2$, $10^3$ and $10^4$ (recall that $C_V \propto T^{d/2}$ for $T \leq T_c$ in $d$ dimensions [32]). Furthermore, smooth curves of $C_V$ vs $T$ with no sign of a phase transition and that exhibit a maximum just above $T_c$ have been observed in path-integral simulations of 2D superfluidity [33] as well as in 2D planar spin models of superfluidity [34]. In those systems, the critical temperature marks the separation between a phase with quasi-long-range order and a disordered one. The crossover from one phase to the other as function of the temperature is known as the Berezinskii-Kosterlitz-Thouless phase transition [35, 36] and differs from the BEC phase transition in that no long-range order is present in the condensate.

Hence, the temperature dependence of the specific heat shown in Fig. 7 (top) over a well-defined range of $T$ may reflect a *dimensional crossover* from 3D to 2D as the strength $\Lambda$ of the $\delta$-barriers is increased. For high temperatures the system undergoes, as expected, another dimensional crossover taking the system back to a 3D behavior. However, this 2D→3D crossover occurs at a characteristic temperature dictated by the relation $\lambda \simeq 0.7a$ [15].

### E. Entropy

The entropy $S$ follows on substituting (6) in the second equation of (9). After some algebra we arrive at the dimensionless entropy per particle

$$\frac{S}{Nk_B} = \frac{1}{N}\ln[N_0 + 1] + \frac{\beta N_0}{N}(\varepsilon_0 - \mu)$$
$$- A\int_0^\infty a dk_z(\tilde{\varepsilon}_{k_z} - \tilde{\mu})\ln\left(1 - \exp\{-\tilde{\beta}(\tilde{\varepsilon}_{k_z} - \tilde{\mu})\}\right)$$
$$+ \frac{2A}{\tilde{\beta}}\int_0^\infty a dk_z g_2(\exp\{-\tilde{\beta}(\tilde{\varepsilon}_{k_z} - \tilde{\mu})\}). \quad (23)$$

The first two terms on the rhs are zero when $T > T_c$ since $N_0 \approx 0$ in the thermodynamic limit $N \longrightarrow \infty$. Thus, the entropy can ultimately be rewritten in terms of sums as

$$\frac{S}{Nk_B} = A\sum_{j=1}^J \int_0^\infty a dk_z(\tilde{\varepsilon}_{k_z} - \tilde{\mu})\ln[1 - \exp\{-\tilde{\beta}(\tilde{\varepsilon}_{k_z} - \tilde{\mu})\}]$$
$$+ \frac{2A}{\tilde{\beta}}\sum_{j=1}^J \int_0^\infty a dk_z g_2(\exp\{-\tilde{\beta}(\tilde{\varepsilon}_{k_z} - \tilde{\mu})\}). \quad (24)$$

Entropy and specific heat curves for $\gamma = 10$ clearly show a smooth transition from a 3D to a 2D behavior for sufficiently large impenetrability $P$ and again to 3D over a finite temperature interval for high enough temperatures. The temperature interval where the system shows a 2D behavior is $P$-dependent and coincides with the specific heat valley in the lower panel in Fig. 7 where $C_V/Nk_B \simeq 1$. The dimensional transition in the entropy is shown as a change in the slope of the $S$ $vs \ln T$ curves of Fig. 8 from $3/2$ in 3D to 1 in 2D. For $\gamma = 0.1$ the dimensional transition is observed at low temperatures and is associated to the changeover of the temperature dependence $T^{3/2} \to T$ (see inset in Fig. 8).

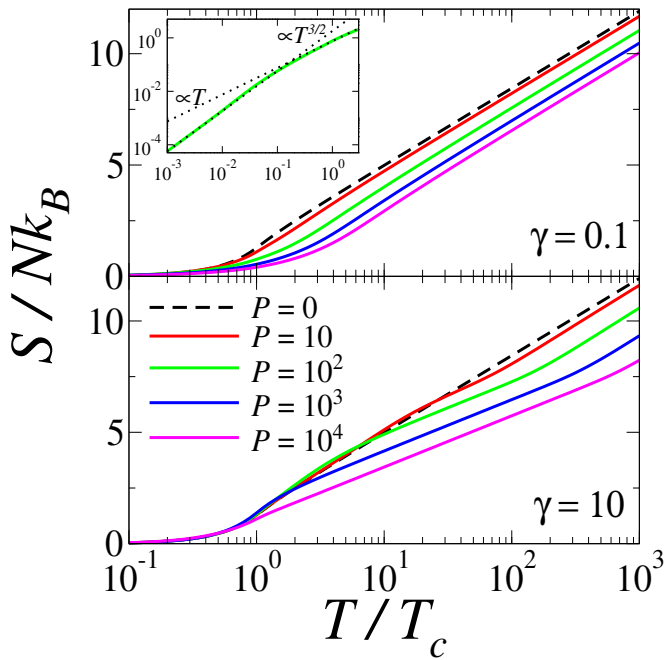

FIG. 8: (Color online) Entropy in $Nk_B$ units as a function of $T/T_c$, for different values of the impenetrability parameter $P$. Upper panel is for $a/\lambda_0 = 0.892$; lower one for $a/\lambda_0 = 0.089$. Inset in the upper panel shows the trapping effect as a change in the slope below the critical temperature.

## III. DENSITY OF STATES

The density of states (DOS) for this system may be obtained as a function of the particle energy $\varepsilon$ and the impenetrability $P$. If $\hbar^2\alpha^2(k_z)/2m$ is the solution of the implicit equation (2), the DOS is

$$
\begin{aligned}
g(\varepsilon) &= \frac{L^2 a}{(2\pi)^3} \int d^3k \, \delta\left(\varepsilon - \frac{\hbar^2}{2m}[k_\perp^2 + \alpha^2(k_z)]\right) \\
&= \frac{L^2 a}{(2\pi)^2} \int dk_\perp k_\perp \int dk_z \, \delta\left(\varepsilon - \frac{\hbar^2}{2m}[k_\perp^2 + \alpha^2(k_z)]\right)
\end{aligned}
$$

where $k_\perp^2 \equiv k_x^2 + k_y^2$. Using cylindrical coordinates and defining $u = \hbar^2 k_\perp^2/2m$ we have

$$
\begin{aligned}
g(\varepsilon) &= \frac{L^2 a}{(2\pi)^2} \frac{m}{\hbar^2} \sum_{j=1}^{\infty} \int_{-\pi/a}^{\pi/a} dk_z \\
&\quad \times \int_0^{\infty} du \, \delta\left(\varepsilon - u - \frac{\hbar^2}{2m}\alpha_j^2(k_z)\right)
\end{aligned}
$$

where $j$ indexes the energy bands in the $z$ direction. Note that the integral over $u$ vanishes if $\varepsilon - \hbar^2\alpha^2/2m < 0$ and equals 1 when $\varepsilon - \hbar^2\alpha^2/2m \geq 0$, or

$$
\int_0^{\infty} du \, \delta\left(u - \left[\varepsilon - \frac{\hbar^2}{2m}\alpha_j^2(k_z)\right]\right) = \theta\left[\varepsilon - \frac{\hbar^2}{2m}\alpha_j^2(k_z)\right]
$$

where $\theta[x]$ is the Heaviside step function. Thus

$$
g(\varepsilon) = \frac{L^2 a}{(2\pi)^2} \frac{m}{\hbar^2} \sum_{j=1}^{\infty} \int_{-\pi/a}^{\pi/a} dk_z \theta\left(\varepsilon - \varepsilon_{k_z}\right). \quad (25)
$$

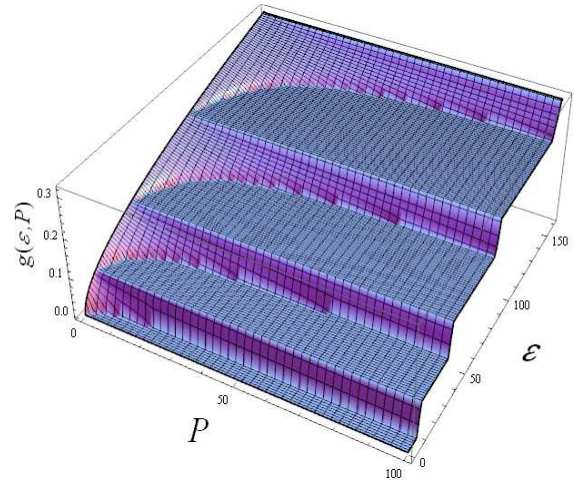

FIG. 9: (Color online) Density of states per particle in units of $\hbar^2/2ma^2$ as a function of the energy $\varepsilon$ in units of $(2ma^2/\hbar^2)$ and impenetrability parameter $P$.

In Fig. 9 $g(\varepsilon)$ is plotted as a function of $\varepsilon$ and the plane impenetrability $P$. For $P = 0$ there are no planes and we recover the 3D DOS $g(\varepsilon) \propto \varepsilon^{1/2}$. For finite $P$ the DOS shows the expected step-like energy dependence. [37] Expression (25) is of little use due to the fact that the sum can not be reduced to an elementary function, however the detailed step-like structure of (25) may be effectively replaced by a two parameters dependent one step structure

$$
g(\varepsilon) = \begin{cases} G \, \varepsilon^{1/2} & \text{if} \quad 0 \leq \varepsilon \leq \varepsilon_{min}, \\ G \, \varepsilon_{min}^{1/2} & \text{if} \quad \varepsilon_{min} < \varepsilon \leq \varepsilon_{min} + \Delta, \\ G \, (\varepsilon - \Delta)^{1/2} & \text{if} \quad \varepsilon_{min} + \Delta < \varepsilon, \end{cases} \quad (26)
$$

where the parameters $\varepsilon_{min}$, $\Delta$ depend on both $P$ and $a/\lambda_0$. $G$ is chosen to meet the exact expression for $g(\varepsilon)$ in the $P = 0$ case. Expression (26) can account for the $T^{3/2} \to T$ crossover of the low-temperature behavior of $C_V/Nk_BT$. For temperatures lower than $T_c$, the specific heat can be written generally as

$$
C_V/k_B = \frac{1}{(k_BT)^2} \int_0^{\infty} d\varepsilon \, g(\varepsilon) \varepsilon^2 \frac{\exp(\beta\varepsilon)}{[\exp(\beta\varepsilon) - 1]^2} \quad (27)
$$

after substituting (26) into last expression and some algebra we have

$$C_V/k_B = G \ (k_B T)^{3/2} I_1(\varepsilon_{min}/k_B T) +$$
$$G \ \varepsilon_{min}^{1/2} k_B T I_2(\varepsilon_{min}/k_B T, \Delta/k_B T) \quad (28)$$

where $I_1 \equiv \int_0^{\varepsilon_{min}/k_B T} dx \, x^{5/2} \exp(x)/[\exp(x) - 1]^2$, a constant for small enough temperature, and $I_2 \equiv \int_{\varepsilon_{min}/k_B T}^{\varepsilon_{min}+\Delta/k_B T} dx \, x^2 \exp(x)/[\exp(x) - 1]^2$. We have disregarded the third region of (26) since this would contribute only at high temperatures. Depending on ratios $\varepsilon_{min}/k_B T$ and $\Delta/k_B T$, one can see that the $T^{3/2}$ term would dominate over the $T$ term for low temperatures. As the temperature is increased, the term proportional to $T$ will dominate giving rise to the crossover.

## IV. CONCLUSIONS

We have calculated the thermodynamic properties of an ideal boson gas in a planar periodic structure as a model to understand different bosonic systems such as Cooper pairs in cuprate superconductors, excitons in multilayer semiconductors, alkali atoms in optical traps or helium-four atoms in multilayer films. The multilayers in the model are generated via the Kronig-Penney delta potential in one direction while the particles are free in the other two directions. A boson gas inside an infinite box is the starting point since its thermodynamic properties are well-known, e.g., its critical BEC temperature $T_0$ which is used to scale the critical temperature of the layered system.

Introducing planes with finite impenetrability $P$ breaks the translational symmetry and this is reflected in the system thermodynamic properties. For $T \leq T_0$ a BEC transition is always present. The critical temperature decreases as a function of $P$ for a fixed separation $a$ between planes until it vanishes for sufficiently large $P$ when the system becomes an infinite set of independent slabs.

The internal energy $U$ increases monotonically as a function of $T$ until it reaches the $3k_B T/2$ classical limit but at a slower rate than the internal energy of systems with smaller $P$ consistent with the specific heat or the entropy behavior.

The specific heat $C_V$ *vs* $T$ for a given $P$ and plane separation $a$, along with the peak associated with BEC exhibits, at temperatures higher than $T_c$, one minimum and one or two maxima. The existence of a local minimum when the thermal wavelength satisfies $\lambda \cong 2a$ suggests that it is caused by bosonic trapping in the $z$-direction since it resembles a quasi-2D behavior. This effect covers a wider temperature region as the impenetrability $P$ increases. However, the system recovers its 3D behavior for plane separations of the order of $1.43\lambda$. The specific heat dependence over a well-defined temperature range as the strength of the $\delta$-barriers is increased reflects a crossover dimensionality effect from 3D to 2D. For every finite $P$, the specific heat per particle reduces to the expected limit $3k_B/2$ at high temperatures.

**Acknowledgments.** Partial support from UNAM-DGAPA-PAPIIT (Mexico) grants IN114708, IN106908 and IN117010-3 as well as CONACyT (Mexico) grant 104917 is gratefully acknowledged.